\documentclass[aps,prl,reprint,superscriptaddress,nofootinbib]{revtex4-2}

\usepackage{amsmath,amssymb,bm}
\usepackage{graphicx}
\usepackage{mathtools}
\usepackage{xcolor}
\usepackage{hyperref}
\usepackage{siunitx}
\usepackage{tikz}
\usetikzlibrary{positioning}
\usepackage{pgfplots}
\pgfplotsset{compat=1.18}

\newcommand{\RR}{\mathbb{R}}

\newcommand{\argmax}{\operatorname*{arg\,max}}

\newcommand{\rank}{\operatorname{rank}}
\newcommand{\sigmin}{\sigma_{\min}}
\newcommand{\Tr}{\operatorname{Tr}}

\begin{document}

\title{Attractor-Keyed Memory }
\author{Natalia G.\ Berloff}
\affiliation{Department of Applied Mathematics and Theoretical Physics, University of Cambridge, Cambridge CB3 0WA, United Kingdom}
\email{N.G.Berloff@damtp.cam.ac.uk}

\begin{abstract}
Physical selectors (lasers choosing a mode, Ising machines settling
on a ground state, condensates occupying a spin state) produce
high-dimensional signatures at the moment of decision: full field
amplitudes, multimode interference patterns, or scattering
responses. These signatures are richer than the winner's index, yet
they are routinely discarded. We show that when the signatures are
repeatable across trials (stereotyped) and linearly independent
across routes, a single linear decoder compiled from calibration data
maps them to arbitrary payloads, merging selection and memory access
into one event and eliminating the fetch that dominates latency and
energy in sparse routing architectures. The construction requires
one singular value decomposition  of measured device responses, which certifies capability and
bounds worst-case error for any downstream payload before the task is
chosen. Runtime error separates into two independently diagnosable
channels, decoding fidelity (controlled by dictionary conditioning) and routing reliability (controlled by the
margin-to-noise ratio), each with a
distinct physical origin and targeted remedy. We derive the full
error decomposition, give Ising-machine selector constructions, and
validate the predicted scalings on synthetic speckle-signature
simulations across three measurement modalities. No hardware
demonstration exists; we provide a falsifiable four-step experimental
protocol specifying what a first experiment must measure. Whether
real device signatures satisfy stereotypy is the central open
question.
\end{abstract}

\maketitle


Architectures built around discrete selection, including
mixture-of-experts
models~\cite{Fedus2022SwitchTransformers}, neuromorphic
processors~\cite{Sze2017DNNsurvey,Izhikevich2025SpikingManifesto}, and
photonic classifiers~\cite{Shen2017NanophotonicDL,Tait2017SiliconWeightBanks},
perform two operations per input: a router selects which of~$K$ experts to activate, then the associated
$D$-dimensional payload is fetched from digital
memory~\cite{wulf1995hitting,horowitz20141,hennessy2019new,Williams2009Roofline}.
The fetch is the bottleneck: the payload read
dominates both latency and
energy~\cite{horowitz20141,Sze2017DNNsurvey}.
Yet the physics of selection has already produced a rich observable at
the moment of decision.

When a laser selects a mode, a condensate occupies a
state~\cite{Berloff2017PolaritonXY}, or an Ising machine
relaxes toward a ground-state
configuration~\cite{McMahon2016,Honjo2021CIM100k,StroevBerloff2023},
the winning attractor carries a high-dimensional physical
signature far richer than the
winner's index.
This signature is discarded.
We show that a fixed linear decoder can map it to arbitrary data, so that selection and memory access become a
single event. We call this primitive
\emph{attractor-keyed memory} (AKM); retrieval
without a separate memory fetch we term \emph{fetchless lookup}.

The framework rests on one physical assumption and one algebraic
condition.
The assumption is \emph{route-conditioned
repeatability} (stereotypy): conditioned on route~$k$ winning,
the measured signature is approximately the same across inputs.
Given stereotypy, the algebraic condition is that the $K$
calibrated mean signatures must be linearly independent.
When they are, a minimum-norm pseudoinverse
decoder~\cite{Penrose1955} $W = Y\Phi^{+}$ recovers any
desired payload table~$Y$ exactly, where
$\Phi \in \RR^{M_{\mathrm{sig}} \times K}$ collects the $K$
mean signatures of dimension~$M_{\mathrm{sig}}$.
Since $\Phi$ depends only on the
hardware, changing the payload requires only recompiling~$W$.

For hardware used as a selector, AKM adds a design objective beyond producing a reliable winner: engineer the post-selection state so its signatures form a well-conditioned dictionary.
This added objective yields three practical consequences. First, a \emph{task-independent
certification protocol}: a single singular value decomposition (SVD) of measured device
responses determines $\rank(\Phi)$ and
$\sigmin(\Phi)$, certifying universal payload
realizability and bounding worst-case error before deployment. Second, a
two-channel error decomposition with distinct diagnostics and
remedies: decoding fidelity controlled by
$\sigmin(\Phi)$, routing reliability controlled by the
ratio $\Delta/T_{\mathrm{eff}}$ of the selection margin to
effective noise temperature. Third, concrete hardware design
targets: three predeployment failure modes (rank loss,
conditioning collapse, margin collapse), each measurable and separately remediable.

The linear algebra is
standard~\cite{Penrose1955}; the rank criterion for~$\Phi$ is
the finite-dictionary analogue of the full-rank condition in
reservoir readout
training~\cite{Jaeger2004,Dambre2012IPC}.
In reservoir computing~\cite{Brunner2013,Dambre2012IPC},
errors fold into a single empirical error budget; in Hopfield
retrieval~\cite{Ramsauer2021}, the attractor \emph{is} the
stored object.
The novelty is the object the algebra acts on:
$\Phi$ is compiled once from measured device responses and
predicts capability for any payload.
Timing-based address selection in spiking
networks~\cite{Berloff2026PWC} and driven-dissipative mode
competition~\cite{Berloff2017PolaritonXY,StroevBerloff2023}
provide candidate physical selectors.

\begin{figure*}[t]
\centering
\includegraphics[width=\textwidth]{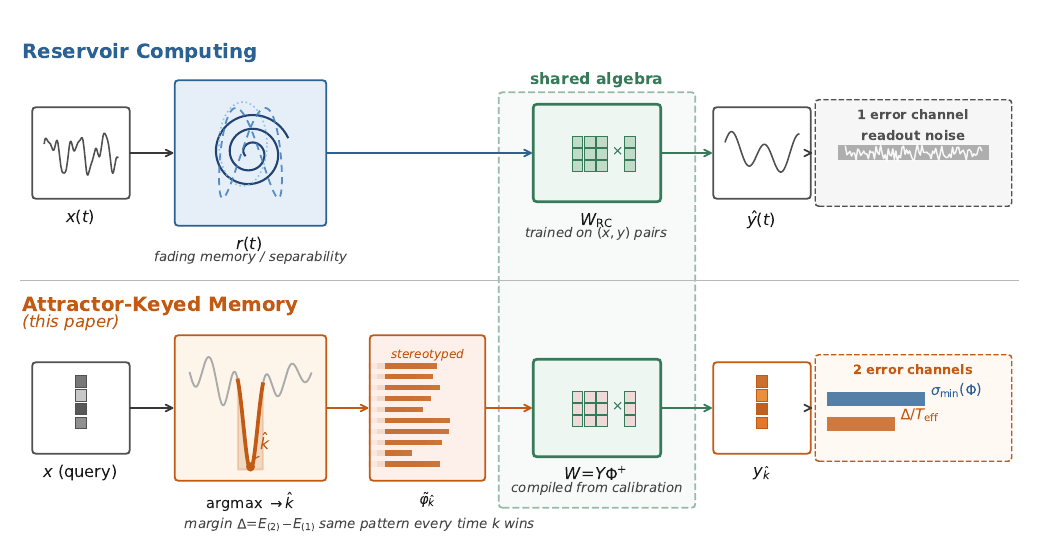}
\caption{
\textbf{Attractor-keyed memory versus reservoir computing.}
Both use a linear readout from a physical state, but differ
structurally: reservoirs operate on a \emph{continuous} driven
trajectory, AKM on a \emph{discrete} set of attractor signatures;
reservoir states are \emph{input-sensitive} (separability), AKM
signatures are \emph{stereotyped} (same pattern each time
route~$k$ wins); the reservoir decoder is \emph{trained} on
input--output pairs, the AKM decoder is \emph{compiled} from
calibration data via $W = Y\Phi^{+}$; reservoir analyses fold
routing-like failures into an overall readout error budget, AKM
decomposes into \emph{two} separately diagnosable channels
(decoding fidelity~$\sigmin(\Phi)$, routing
reliability~$\Delta/T_{\mathrm{eff}}$); reservoir analysis
yields a capacity bound, AKM yields a predeployment
\emph{certification protocol}.
In AKM the payload~$y_k$ need not resemble the attractor state;
the attractor is the key, not the stored value.
}
\label{fig:akm_vs_rc}
\end{figure*}

\begin{figure*}[t!h!]
\centering
\includegraphics[width=0.85\textwidth]{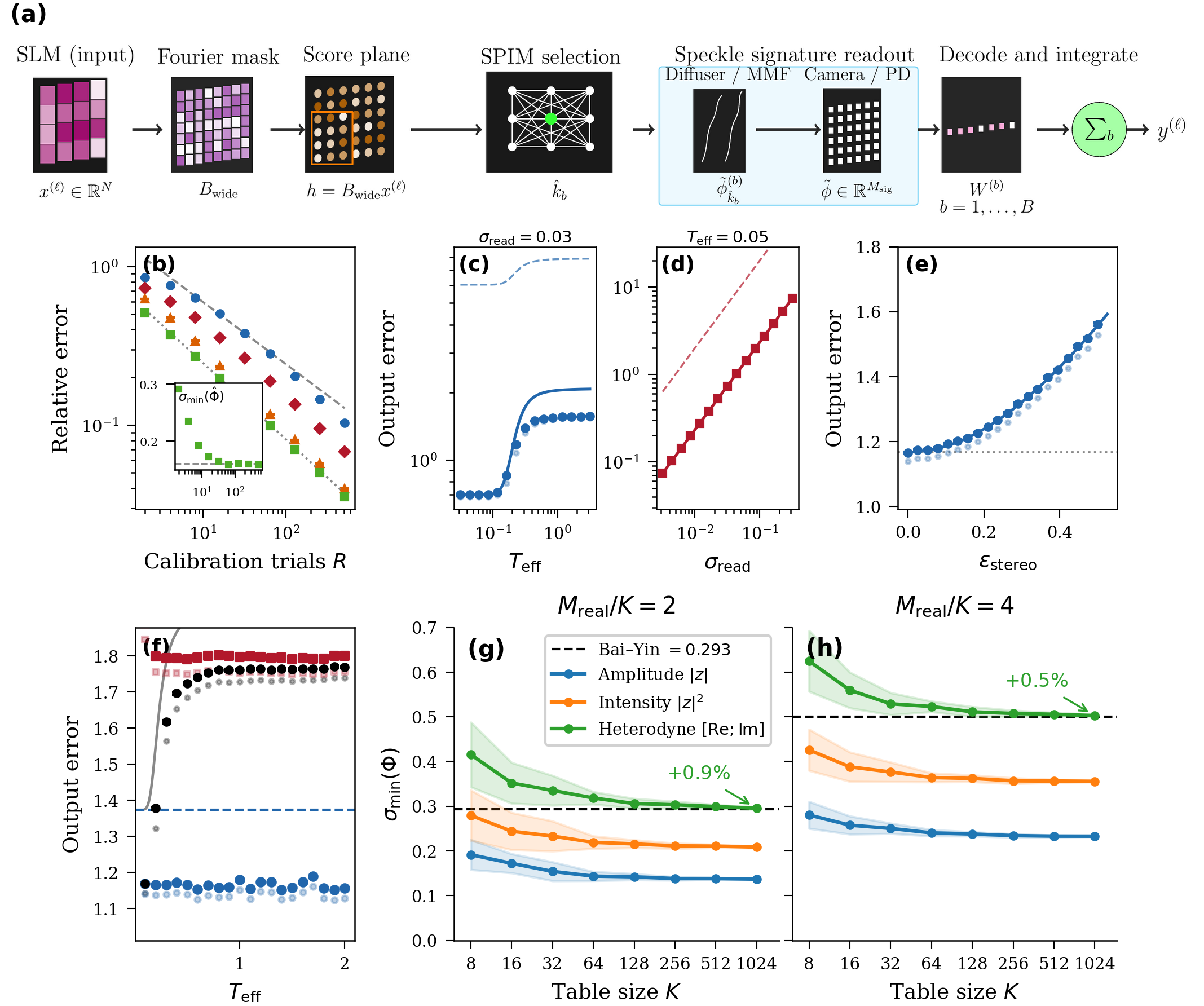}
\caption{
\textbf{Photonic implementation and numerical validation of
attractor-keyed memory.}
\textbf{(a)}~Schematic photonic pipeline.
A spatial light modulator (SLM) encodes the input~$x$; a Fourier
mask implements~$B_{\mathrm{wide}}$.
A routing map~$R_{\mathrm{route}}$ compresses scores
into~$K$ competing routes. A spatial photonic Ising machine
(SPIM)~\cite{Pierangeli2019PhotonicIsing_PRL} selects the
winner~$\hat{k}_b$ with margin~$\Delta$. The selected mode's
field propagates through a scattering medium; a detector array records the speckle
signature~$\tilde\phi_{\hat{k}} \in \RR^{M_{\mathrm{sig}}}$.
A pre-compiled decoder~$W^{(b)}$ maps each signature to output;
summing over~$B$ blocks yields~$y$.
\textbf{(b)--(f)}~Error decomposition on a controlled
speckle-signature model (Monte Carlo, not experiment;
$K\!=\!64$, $M_{\mathrm{sig}}\!=\!128$, $D\!=\!16$;
$R$: calibration trials per route;
$\sigma_{\mathrm{read}}$: readout noise s.d.).
\textbf{(b)}~Relative reconstruction error
$\|W_{\hat\Phi}\,\Phi - Y\|_F/\|Y\|_F$ versus calibration trials~$R$;
four dictionary types (amplitude~$|z|$,
intensity~$|z|^2$, heterodyne~$[\Re\,z;\Im\,z]$,
orthogonal); $\sigmin(\Phi)$ controls convergence rate.
\textbf{(c)--(f)}~All four panels use the amplitude-speckle
dictionary.
Solid lines: analytic RMS prediction
$\sqrt{\mathbb{E}\|e\|_2^2}$ from the two-channel
second-moment decomposition (Supp.\ Mat., Remark~3,
Eq.~(S17));
filled circles: Monte Carlo RMS ($10{,}000$ trials);
faint circles: Monte Carlo mean~$\mathbb{E}[\|e\|_2]$;
dashed: per-trial worst-case bounds
[Eq.~\eqref{eq:driftbound} for decoding;
$d_Y + \|Y\|_2\,\bar\delta/\sigmin(\Phi)$
for misrouting] weighted by $P_{\mathrm{mis}}$
from Eq.~\eqref{eq:misbound}.
\textbf{(c)}~Routing-dominated regime (varying~$T_{\mathrm{eff}}$
at fixed $\sigma_{\mathrm{read}}\!=\!0.03$).
\textbf{(d)}~Decoding-dominated regime (varying
$\sigma_{\mathrm{read}}$ at fixed $T_{\mathrm{eff}}\!=\!0.05$).
The two error channels are separable and each matches its
predicted scaling.
\textbf{(e)}~Decoding error under stereotypy violation
$\varepsilon_{\mathrm{stereo}}$ (correct routing enforced).
\textbf{(f)}~Full pipeline at a realistic operating point;
the crossover at $T_{\mathrm{eff}}\!\approx\!0.71$ separates
the decoding-limited from the routing-limited regime.
\textbf{(g),\,(h)}~Dictionary conditioning~$\sigmin(\Phi)$
versus table size~$K$ at fixed aspect ratio
$M_{\mathrm{real}}/K$.
Three measurement modalities,
30 random seeds per point; shaded bands: $\pm 1$~s.d.
\textbf{(g)}~$M_{\mathrm{real}}/K\!=\!2$.
\textbf{(h)}~$M_{\mathrm{real}}/K\!=\!4$.
From $K\!=\!64$ onward, $\sigmin$ tracks the Bai--Yin
prediction (dashed). At $K\!=\!1024$ the heterodyne value lies within
$1\%$ of the asymptote at both aspect ratios.
See Supp.\ Mat.~\cite{SM} for full panel specifications.
}
\label{fig:combined}
\end{figure*}

\emph{Evidence hierarchy.}
Four layers, in decreasing rigor:
(i)~exact theorem (full-rank dictionary gives
universal payload realizability); (ii)~approximate theory
(conditioning, drift, and stereotypy bounds);
(iii)~phenomenology (Gibbs routing fit); (iv)~synthetic
validation (Monte Carlo on speckle-signature models, not
physical experiment).
Synthetic validation confirms internal consistency; it does not test physical assumptions.
Figure~\ref{fig:akm_vs_rc} contrasts the
architecture with reservoir computing;
Fig.~\ref{fig:combined}(a) illustrates a photonic
realization.
Whether real competitive selectors satisfy stereotypy is the central
open experimental question; the formal statement and its quantitative
relaxation appear below.


\paragraph{Route--decode abstraction.}
Fetchless lookup factors into routing and readout:
\begin{equation}
 x \;\longmapsto\; \hat{k}(x)
 \;\longmapsto\; \tilde\phi_{\hat{k}(x)}
 \;\longmapsto\; y = W\tilde\phi_{\hat{k}(x)}.
\label{eq:route_decode}
\end{equation}
Here $x \in \RR^N$ is an $N$-dimensional input; $\hat{k}(x)$ is
the discrete route selected by the physical competition;
$\tilde\phi_{\hat{k}(x)} \in \RR^{M_{\mathrm{sig}}}$ is the
single-shot measured signature of the winning state; and
$W \in \RR^{D \times M_{\mathrm{sig}}}$ is a fixed linear decoder
mapping signatures to $D$-dimensional payloads.
The theory requires only that routing returns a
winner~$\hat{k}(x)$ with a measurable margin; the internal structure of the
selector is an implementation detail.
In these terms, a \emph{route} is the index~$\hat{k}(x)$ returned by the
selector; the \emph{payload}~$y_k$ is the data assigned to route~$k$; a
\emph{fetch} is the separate memory read that returns~$y_k$ from a stored table
once the route is known; and \emph{fetchless lookup} eliminates that read by
obtaining~$y$ directly from the winner's signature through~$W$, as in
Eq.~\eqref{eq:route_decode}.


\paragraph{Score generation and routing.}
One realization maps an input~$x$ to a bank of
$M_{\mathrm{score}} \ge K$ candidate scores:
\begin{equation}
 h(x) = B_{\mathrm{wide}}\,x,
 \qquad
 B_{\mathrm{wide}}\in\RR^{M_{\mathrm{score}}\times N}.
\label{eq:wide}
\end{equation}
In photonic platforms $B_{\mathrm{wide}}$ may be realized by a
programmable linear optical
transformation~\cite{Miller2013,Clements2016,Onodera2025}.
A fixed routing map
$R_{\mathrm{route}} \in \RR^{K \times M_{\mathrm{score}}}$
compresses the wide scores into $K$ competing routes:
\begin{equation}
 g(x) = R_{\mathrm{route}}\,h(x)\in\RR^{K},
 \qquad
 E_k(x) = -g_k(x),
\label{eq:block}
\end{equation}
where $g_k(x)$ is the score for route~$k$ and $E_k(x)$ the
corresponding selector energy. The selector returns
$\hat{k}(x) \in \argmax_k g_k(x)$ with winner-to-runner-up
margin
\begin{equation}
 \Delta(x) = g_{(1)}(x) - g_{(2)}(x),
\label{eq:selector_gap}
\end{equation}
where $g_{(1)} \ge g_{(2)} \ge \cdots$ are the ordered scores.
Since $E_k = -g_k$, this score-space margin equals the
energy-space gap $E_{(2)} - E_{(1)}$; we use $\Delta$ for both
throughout.
Equations~\eqref{eq:wide}--\eqref{eq:selector_gap} describe one
realization; the theory requires only a winner~$\hat{k}(x)$,
a margin~$\Delta(x)$, and the testable assumption that
misrouting decreases monotonically with~$\Delta(x)$.


\paragraph{Signature emission and decoding.}
After routing converges to state~$u^{(k)}$, a fixed measurement
map~$\mathcal{M}$ produces the signature
\begin{equation}
 \tilde\phi_k = \mathcal{M}\,u^{(k)} + \delta
 \in\RR^{M_{\mathrm{sig}}},
\label{eq:signature}
\end{equation}
where $\delta$ is zero-mean measurement noise
(decomposed in Supp.\ Mat., Sec.~S3D into an input-dependent
residual~$\varepsilon_k(x)$ and a stochastic
component~$\delta_k$).
Conditioned on selecting route~$k$, the measured signature has
mean~$\bar\phi_k$ and covariance~$\Sigma_k$.

\emph{Assumption (route-conditioned repeatability / stereotypy).}
Conditioned on route~$k$ winning, shot-to-shot variation
in~$\tilde\phi_k$ is dominated by device noise, not by
differences in the triggering input~$x$.
Strong competition funnels all initial conditions toward
a single final state; weak competition lets the winning
state retain input memory, making $\Phi$ a poor summary.
Mode hopping or near-degeneracy can break
stereotypy and must be diagnosed (Supp.\ Mat., Sec.~S3).

Calibration forces each route in turn and collects mean signatures and payloads:
\begin{equation}
 \Phi = [\bar\phi_1\cdots\bar\phi_K]
 \in\RR^{M_{\mathrm{sig}}\times K},
 \qquad
 Y = [y_1\cdots y_K]\in\RR^{D\times K}.
\nonumber
\end{equation}
The dictionary~$\Phi$ is empirical, compiled from measured device
responses. Runtime readout uses a fixed linear decoder
$
 y = W\tilde\phi_{\hat{k}(x)},
$ with $
 W\in\RR^{D\times M_{\mathrm{sig}}}.
$


\paragraph{Block-parallel architecture.}
$B$ independent blocks operate in parallel, each running
its own physical $\argmax$ over $K$ routes; the rank condition
applies per block. The layer output is the sum of decoded
signatures across all $B$ blocks (Supp.\ Mat.~\cite{SM}).
Figure~\ref{fig:combined}(a) illustrates the pipeline.


\paragraph*{Proposition 1 (Universal payload realizability).}
The system $W\Phi = Y$ admits a solution for every
$Y\in\RR^{D\times K}$ if and only if $\rank(\Phi)=K$.
When this holds, $M_{\mathrm{sig}}\ge K$ and the
minimum-norm exact decoder is $W_\star=Y\Phi^{+}$
(proof and full decoder family in Supp.\ Mat.~\cite{SM}).
The condition is well known; what matters here is its
\emph{physical interpretation}: the $K$ routes must produce $K$
linearly independent measured patterns, and a single SVD of~$\Phi$
checks this before any task is specified.

If $\rank(\Phi) = r < K$, exact decoding may still hold for a
particular~$Y$ whose rows lie in $\operatorname{Row}(\Phi)$. Rank
deficiency rules out \emph{universal} fetchless lookup, not every
structured task.


\paragraph{Selector realizations.}
The decoder theory is selector-agnostic, but two Ising
constructions show the framework generates concrete hardware
designs.

\emph{One-hot QUBO (quadratic unconstrained binary optimization)
selector.}
Binary variables $z_i \in \{0,1\}$ define the energy
\begin{equation}
 E_{\mathrm{WTA}}(z;x)
 = \lambda\Big(\sum_{i=1}^{K}z_i - 1\Big)^{\!2}
 - \sum_{i=1}^{K}g_i(x)\,z_i,
 \qquad \lambda>0,
\label{eq:wta_qubo}
\end{equation}
where $\lambda$ is a penalty weight enforcing the one-hot
constraint. If $\lambda$ exceeds the largest score magnitude,
every global minimizer is one-hot: on the one-hot manifold,
$E_{\mathrm{WTA}}(e_i;x) = -g_i(x)$, recovering the selector
energy of Eq.~\eqref{eq:block}. The standard spin transformation
$z_i = (1+s_i)/2$ yields an equivalent Ising Hamiltonian with
dense antiferromagnetic couplings and local fields proportional
to~$g_i(x)$. Each block in the parallel architecture realizes
its own such QUBO. The full proof is
given in the Supp.\ Mat.~\cite{SM}.

\emph{Binary comparator ($K\!=\!2$).}
An antiferromagnetically coupled spin pair with coupling
$J > 0$ and local fields $h_a(x), h_b(x)$ returns
$\operatorname{sign}(h_a - h_b)$~\cite{Izhikevich2025SpikingManifesto};
when $J$ exceeds the field magnitudes, the energy gap is
$\Delta_{\mathrm{cmp}}(x) = 2|h_a(x) - h_b(x)|$. Chaining
$n_c$ such comparators produces an $n_c$-bit address into a
$2^{n_c}$-entry lookup table. Sweeping the field difference and
comparing the empirical misrouting rate against
$\Delta_{\mathrm{cmp}}/T_{\mathrm{eff}}$ is the simplest
experimental test of fetchless lookup.

Driven-dissipative oscillators offer a third realization
(Supp.\ Mat.~\cite{SM}).


\paragraph{Robustness: two separable failure channels.}
At run time two error channels remain: signature perturbation
after the correct route wins, and selection of the wrong route.
The decomposition is a diagnostic tool, not a claim of
statistical independence (Supp.\ Mat.~\cite{SM}, Remark~3).

\emph{Conditional decoding error.}
If route~$k$ wins and the single-shot signature is
$\tilde\phi_k = \bar\phi_k + \delta$ (with perturbation $\delta$
aggregating shot noise, detector noise, and drift), the
minimum-norm decoder gives (using $W\bar\phi_k = y_k$)
\begin{equation}
 \|W\tilde\phi_k - y_k\|_2
 = \|Y\Phi^{+}\delta\|_2
 \le \frac{\|Y\|_2}{\sigmin(\Phi)}\,\|\delta\|_2,
\label{eq:driftbound}
\end{equation}
where $\|Y\|_2$ is the spectral norm of the payload table. For
zero-mean fluctuations with covariance~$\Sigma_k$,
\begin{equation}
 \mathbb{E}\|W\tilde\phi_k-y_k\|_2^2
 = \Tr\!\big[Y\Phi^{+}\Sigma_k(\Phi^{+})^{\!\top}Y^{\!\top}\big]
 \le \frac{\|Y\|_2^2}{\sigmin(\Phi)^2}\,\Tr\Sigma_k.
\label{eq:msebound}
\end{equation}

When stereotypy is imperfect and the residual input-dependent
shift has norm at most~$\varepsilon_{\mathrm{stereo}}$, the
additional decoding error is bounded by
$\|Y\|_2\,\varepsilon_{\mathrm{stereo}}/\sigmin(\Phi)$
(Supp.\ Mat.~\cite{SM}), so stereotypy need only hold to within
the device noise floor.

\emph{Dictionary drift.}
If the dictionary drifts from~$\Phi_0$ at calibration to
$\Phi_0 + \delta\!\Phi(t)$ while the decoder remains
$W_0 = Y\Phi_0^{+}$,
\begin{equation}
 \|W_0[\Phi_0+\delta\!\Phi(t)]-Y\|_2
 \le \frac{\|Y\|_2}{\sigmin(\Phi_0)}\,\|\delta\!\Phi(t)\|_2.
\label{eq:dictdrift}
\end{equation}
Recalibration is needed once the drift exceeds tolerance
$\varepsilon_{\mathrm{tol}}\,\sigmin(\Phi_0)$.
Each recalibration costs $O(KR)$ forced-route measurements ($R$ trials per
route), so the duty-cycle overhead is this cost divided by the drift interval
$\tau_{\mathrm{drift}}$ over which $\|\delta\!\Phi(t)\|_2$ stays within
tolerance. Since $\tau_{\mathrm{drift}}$ is device-set and presently
uncharacterised, the overhead cannot be quantified without a drift measurement
on a specific platform; the experimental protocol below measures it directly.

\emph{Routing error.}
If the intended route is~$k^\star$ but the selector returns
$\ell \neq k^\star$, no decoder can repair the
mistake. We use the Gibbs form as a phenomenological fit; the
theory needs only that larger margin means lower misrouting:
\begin{equation}
 P(k) = \frac{e^{-E_k/T_{\mathrm{eff}}}}
 {\sum_j e^{-E_j/T_{\mathrm{eff}}}},
\label{eq:gibbs}
\end{equation}
where $E_k(x) = -g_k(x)$ is the selector energy defined in
Eq.~\eqref{eq:block} and $T_{\mathrm{eff}}$ is an effective
temperature fitted from repeated
trials~\cite{McMahon2016,Hamerly2019,StroevBerloff2023},
the misrouting probability satisfies
\begin{equation}
 P_{\mathrm{mis}}
 \le \frac{(K-1)\,e^{-\Delta/T_{\mathrm{eff}}}}
 {1+(K-1)\,e^{-\Delta/T_{\mathrm{eff}}}},
\label{eq:misbound}
\end{equation}
where $\Delta = \min_{k \neq k^\star}(E_k - E_{k^\star})$ is the
minimum energy gap to the next competing route
(equal to the score-space margin of
Eq.~\eqref{eq:selector_gap}, since $E_k = -g_k$). The bound follows from bounding each competitor's
Boltzmann weight by $e^{-\Delta/T_{\mathrm{eff}}}$;
for $\Delta \gg T_{\mathrm{eff}}$, log-odds of correct
routing scale linearly in $\Delta/T_{\mathrm{eff}}$.
Near-degeneracy, where route statistics in Ising machines and SPIMs are known
to depart from Boltzmann behaviour, is where the Gibbs form is least reliable;
there the structural results (the two-channel decomposition and the
certification) still hold, as they assume only monotonicity, while the specific
bound~\eqref{eq:misbound} should be replaced by the empirically measured
misrouting-versus-margin curve.
With payload diameter
$d_Y = \max_{k,\ell}\|y_k - y_\ell\|_2$, the triangle
inequality gives a routing contribution at most
$P_{\mathrm{mis}}\,(d_Y
+ \|Y\|_2\,\bar\delta/\sigmin(\Phi))$,
where $\bar\delta = \max_k\sqrt{\Tr\Sigma_k}$ is the worst-case
per-route noise level.
A tighter second-moment decomposition is given in the Supp.\ Mat.
(Remark~3); it requires the
additional modeling assumption that route-conditioned emission
noise has zero mean and covariance~$\Sigma_\ell$ on each route
$\ell$, independently of which route was intended (see Supp.\ Mat.
for the precise statement).
Figure~\ref{fig:combined}(b--f) illustrates the decomposition on a
controlled speckle-signature model (Monte Carlo;
see Supp.\ Mat., Fig.~S1).
Because the simulation draws routes from the same Gibbs
model used to derive Eq.~\eqref{eq:misbound}, the agreement
confirms the algebra, not the
physical adequacy of the Gibbs assumption;
that requires a goodness-of-fit test on real route
frequencies, not yet available.


\paragraph{Calibration and training.}
Fetchless lookup operates on two time scales. A slow calibration
stage forces each route, measures signature statistics,
forms~$\Phi$, and compiles $W = Y\Phi^{+}$. Between
recalibrations, $\Phi$ is fixed and online learning updates only~$Y$
(one column per sample, rank-1 update in~$W$);
backpropagation through the $\argmax$ uses a top-two surrogate gradient
(Supp.\ Mat.~\cite{SM}).


\paragraph{Experimental protocol and falsifiability.}
Four steps translate the theory into a falsifiable experiment:
{\bf(1)}
Force each route in turn and record $R$ repeated
 signature measurements.
 This determines the sample-mean dictionary~$\hat\Phi$
 (the finite-sample estimate of~$\Phi$),
 the covariances~$\Sigma_k$, the
 empirical rank, $\sigmin(\hat\Phi)$, and the stereotypy diagnostic $\varepsilon_{\mathrm{stereo}}/\!\sqrt{\Tr\Sigma_k}$
 (Supp.\ Mat., Sec.~S3). Full rank is confirmed only if a bootstrap confidence
interval for $\sigmin(\hat\Phi)$ excludes zero;
the interval's lower endpoint provides a conservative
bound on dictionary conditioning. Under the hypothesis that
centered signatures are sub-Gaussian with
$\sigma_{\mathrm{sg},k}^2 \le \|\Sigma_k\|_2$, the number of trials needed to resolve
$\sigmin(\Phi)$ scales as
$R = O(K\,\|\Sigma\|_2\,(M_{\mathrm{sig}} + \log K)/\sigmin(\Phi)^2)$
(Supp.\ Mat., Proposition~2). If $\rank(\Phi) < K$,
no linear decoder can realize universal fetchless lookup.
 {\bf (2)} Compile $W = Y\Phi^{+}$ for test payloads and verify
 $W\Phi \approx Y$ on forced-route means; compare single-shot
 error scaling with $\sigmin(\Phi)$ via
 Eqs.~\eqref{eq:driftbound}--\eqref{eq:msebound}.
 {\bf (3)} Release the selector, record winner frequencies versus
 the measured top-two margin, and fit~$T_{\mathrm{eff}}$.
 Assess the Gibbs fit by a goodness-of-fit test
 (e.g., $\chi^2$ on binned route frequencies); test
 Eq.~\eqref{eq:misbound}.
 A necessary consistency check: if forced-route and
 free-running signatures differ significantly, the
 calibration model requires correction.
 {\bf (4)} Monitor drift in mean signatures over time.
 Recalibrate when $\|\delta\!\Phi(t)\|_2$ exceeds
$\varepsilon_{\mathrm{tol}}\,\sigmin(\Phi_0)$,
 the tolerance derived from Eq.~\eqref{eq:dictdrift}.

No hardware demonstration exists to date; the
protocol defines the criteria a first experiment must satisfy.
The required ingredients exist separately in current platforms~\cite{Brunner2013,Hughes2019,Miller2013,Clements2016,Onodera2025,Pai2023,McMahon2016,Kalinin2018,StroevBerloff2023}; integrating them into a
single device remains open.
A forced-routing
dictionary measurement is the natural first experiment, followed
by a $K = 2$ routing test of the predicted
$\Delta_{\mathrm{cmp}}/T_{\mathrm{eff}}$ dependence.

\paragraph{Scope and outlook.}
How far the scheme scales depends on the oversampling ratio
$M_{\mathrm{sig}}/K$. The Bai--Yin law keeps
$\sigmin(\Phi)$ bounded away from zero when
$M_{\mathrm{sig}}/K$ exceeds unity by a finite factor,
yielding low error amplification at
$M_{\mathrm{sig}}/K = 2$
[Eqs.~\eqref{eq:driftbound}--\eqref{eq:msebound}].
Real device signatures may exhibit spatial correlations that
reduce the effective degrees of freedom below $M_{\mathrm{sig}}$,
worsening conditioning; the SVD diagnostic detects this.
Among the three readout modalities tested
[Fig.~\ref{fig:combined}(b) and (g,\,h)],
heterodyne achieves the best conditioning, improving $\sigmin$
by $3.4\times$ over amplitude-only measurement at matched
complex-mode count, and remains within $1\%$ of the Bai--Yin
asymptote up to $K\!=\!1024$ (Supp.\ Mat., Fig.~S2).

AKM replaces an $O(D)$ memory read with an
$M_{\mathrm{sig}}$-channel measurement and linear decode, favourable when the decode is absorbed into the measurement
optics or when data-movement cost dominates
compute~\cite{horowitz20141,Sze2017DNNsurvey}; at $M_{\mathrm{sig}}/K = 2$ with digital decode, break-even
requires native optical decode or DRAM-resident payloads
(Supp.\ Mat., Sec.~S13).

Any competitive physical selector that produces
high-dimensional signatures and settles to stereotyped attractor
states is a candidate platform: coherent Ising machines,
polariton condensate networks, laser arrays, and spatial photonic
Ising machines (SPIMs). Among these, SPIMs are closest to the
requirements: focal-plane division now enables fully programmable
Ising selection~\cite{Veraldi2025PRL}, and full-aperture wavefront
correction removes the aberration bottleneck that previously
limited effective $M_{\mathrm{sig}}$~\cite{Karanikolopoulos2026}.
What remains untested is the quantity AKM requires:
within-attractor variance of the full speckle signature,
conditioned on the same winning route.

Indirect evidence is encouraging: speckle physical unclonable functions (PUFs), photonic
reservoirs, and polariton condensates achieve
$\gtrsim\!94\%$ reproducibility of attractor-level
observables~\cite{gao2020physical,Pappu2002PUF,Oliver2015Consistency,%
Brunner2013,Ohadi2015SpinBifurcation,Pickup2018SpinFlips,Hamerly2019},
but the \emph{continuous high-dimensional
state} within a given attractor is almost never
quantified~\cite{SM}.
A first experiment need only record continuous-valued output
conditioned on the same attractor, yielding the four quantities $\Phi$,
$\sigmin(\Phi)$, $\Sigma_k$,
and $\varepsilon_{\mathrm{stereo}}$ that
determine whether a platform supports fetchless lookup at a
target scale.
\begin{acknowledgments}
The author acknowledges support from HORIZON EIC-2022-PATHFINDERCHALLENGES-01 HEISINGBERG Project 101114978, from Weizmann--UK Make Connection Grant 142568, and from the EPSRC UK Multidisciplinary Centre for Neuromorphic Computing (grant UKRI982).
\end{acknowledgments}
{\it Patent and Implementation Notice.} Certain systems, methods, hardware configurations, acceleration techniques, implementation architectures, and commercial applications related to the work described in this manuscript are the subject of pending or patent applications in progress. This manuscript is intended to describe the scientific concepts and experimental framework at a research level and does not disclose all proprietary engineering, hardware, system-integration, optimization, or commercial implementation details.

\bibliographystyle{apsrev4-2}
\bibliography{refs2}

@article{McMahon2016,
  author = {McMahon, Peter L. and Marandi, Alireza and Haribara, Yoshitaka and Hamerly, Ryan and Langrock, Carsten and Tamate, Shuhei and Inagaki, Takahiro and Takesue, Hiroki and Utsunomiya, Shoko and Aihara, Kazuyuki and Byer, Robert L. and Fejer, M. M. and Mabuchi, Hideo and Yamamoto, Yoshihisa},
  title = {A fully programmable 100-spin coherent Ising machine with all-to-all connections},
  journal = {Science},
  volume = {354},
  number = {6312},
  pages = {614--617},
  year = {2016},
  doi = {10.1126/science.aah5178}
}

@article{Hamerly2019,
  author = {Hamerly, Ryan and Bernstein, Liane and Sludds, Alexander and Solja{\v{c}}i{\'c}, Marin and Englund, Dirk},
  title = {Experimental investigation of performance differences between coherent Ising machines and a quantum annealer},
  journal = {Physical Review X},
  volume = {9},
  pages = {021032},
  year = {2019},
  doi = {10.1103/PhysRevX.9.021032}
}

@article{Kalinin2018,
  author = {Kalinin, Kirill P. and Berloff, Natalia G.},
  title = {Simulating {Ising} and $n$-State Planar {Potts} Models and External Fields with Nonequilibrium Condensates},
  journal = {Physical Review Letters},
  volume = {121},
  pages = {235302},
  year = {2018},
  doi = {10.1103/PhysRevLett.121.235302}
}

@article{StroevBerloff2023,
  author = {Stroev, Nikita and Berloff, Natalia G.},
  title = {Analog Photonics Computing for Information Processing, Inference, and Optimization},
  journal = {Advanced Quantum Technologies},
  volume = {6},
  number = {9},
  pages = {2300055},
  year = {2023},
  doi = {10.1002/qute.202300055}
}

@article{Miller2013,
  author = {Miller, David A. B.},
  title = {Self-configuring universal linear optical component [Invited]},
  journal = {Photonics Research},
  volume = {1},
  number = {1},
  pages = {1--15},
  year = {2013},
  doi = {10.1364/PRJ.1.000001}
}

@article{Clements2016,
  author = {Clements, William R. and Humphreys, Peter C. and Metcalf, Benjamin J. and Kolthammer, W. Steven and Walmsley, Ian A.},
  title = {Optimal design for universal multiport interferometers},
  journal = {Optica},
  volume = {3},
  number = {12},
  pages = {1460--1465},
  year = {2016},
  doi = {10.1364/OPTICA.3.001460}
}

@article{Brunner2013,
  author = {Brunner, Daniel and Soriano, Miguel C. and Mirasso, Claudio R. and Fischer, Ingo},
  title = {Parallel photonic information processing at gigabyte per second data rates using transient states},
  journal = {Nature Communications},
  volume = {4},
  pages = {1364},
  year = {2013},
  doi = {10.1038/ncomms2368}
}

@article{Hughes2019,
  author = {Hughes, Tyler W. and Williamson, Ian A. D. and Minkov, Momchil and Fan, Shanhui},
  title = {Wave physics as an analog recurrent neural network},
  journal = {Science Advances},
  volume = {5},
  number = {12},
  pages = {eaay6946},
  year = {2019},
  doi = {10.1126/sciadv.aay6946}
}

@article{Pai2023,
  author = {Pai, Sunil and Sun, Zhanghao and Hughes, Tyler W. and Park, Taewon and Bartlett, Ben and Williamson, Ian A. D. and Minkov, Momchil and Milanizadeh, Maziyar and Abebe, Nathnael and Morichetti, Francesco and Melloni, Andrea and Fan, Shanhui and Solgaard, Olav and Miller, David A. B.},
  title = {Experimentally realized in situ backpropagation for deep learning in photonic neural networks},
  journal = {Science},
  volume = {380},
  number = {6643},
  pages = {398--404},
  year = {2023},
  doi = {10.1126/science.ade8450}
}

@article{Onodera2025,
  author = {Onodera, Tatsuhiro and Stein, Martin M. and Ash, Benjamin A. and Sohoni, Mandar M. and Bosch, Melissa and Yanagimoto, Ryotatsu and Jankowski, Marc and McKenna, Timothy P. and Wang, Tianyu and Shvets, Gennady and Shcherbakov, Maxim R. and Wright, Logan G. and McMahon, Peter L.},
  title = {Arbitrary control over multimode wave propagation for machine learning},
  journal = {Nature Physics},
  year = {2025},
  doi = {10.1038/s41567-025-03094-2}
}

@article{Penrose1955,
  author = {Penrose, Roger},
  title = {A generalized inverse for matrices},
  journal = {Mathematical Proceedings of the Cambridge Philosophical Society},
  volume = {51},
  number = {3},
  pages = {406--413},
  year = {1955},
  doi = {10.1017/S0305004100030401}
}

@misc{Izhikevich2025SpikingManifesto,
  title         = {Spiking Manifesto},
  author        = {Izhikevich, Eugene},
  year          = {2025},
  eprint        = {2512.11843},
  archivePrefix = {arXiv},
  primaryClass  = {cs.NE},
  doi           = {10.48550/arXiv.2512.11843},
  url           = {https://arxiv.org/abs/2512.11843}
}

@misc{Berloff2026PWC,
  title         = {Polychronous Wave Computing: Timing-Native Address Selection in Spiking Networks},
  author        = {Berloff, Natalia G.},
  year          = {2026},
  eprint        = {2601.13079},
  archivePrefix = {arXiv},
  primaryClass  = {cond-mat.dis-nn},
  doi           = {10.48550/arXiv.2601.13079},
  url           = {https://arxiv.org/abs/2601.13079}
}

@article{Pierangeli2019PhotonicIsing_PRL,
  title   = {Large-Scale Photonic Ising Machine by Spatial Light Modulation},
  author  = {Pierangeli, Davide and Marcucci, Giulia and Conti, Claudio},
  journal = {Physical Review Letters},
  volume  = {122},
  number  = {21},
  pages   = {213902},
  year    = {2019},
  doi     = {10.1103/PhysRevLett.122.213902}
}

@article{Williams2009Roofline,
  author  = {Williams, Samuel and Waterman, Andrew and Patterson, David A.},
  title   = {Roofline: An Insightful Visual Performance Model for Floating-Point Programs and Multicore Architectures},
  journal = {Communications of the ACM},
  volume  = {52},
  number  = {4},
  pages   = {65--76},
  year    = {2009},
  doi     = {10.1145/1498765.1498785}
}

@article{Sze2017DNNsurvey,
  author  = {Sze, Vivienne and Chen, Yu-Hsin and Yang, Tien-Ju and Emer, Joel S.},
  title   = {Efficient Processing of Deep Neural Networks: A Tutorial and Survey},
  journal = {Proceedings of the IEEE},
  volume  = {105},
  number  = {12},
  pages   = {2295--2329},
  year    = {2017},
  doi     = {10.1109/JPROC.2017.2761740}
}

@article{Shen2017NanophotonicDL,
  author  = {Shen, Yichen and Harris, Nicholas C. and Skirlo, Scott and Prabhu, Mishav M. and Baehr-Jones, Tom and Hochberg, Michael and Sun, Xin and Zhao, Shijie and Larochelle, Hugo and Englund, Dirk and Solja{\v c}i{\'c}, Marin},
  title   = {Deep Learning with Coherent Nanophotonic Circuits},
  journal = {Nature Photonics},
  volume  = {11},
  pages   = {441--446},
  year    = {2017},
  doi     = {10.1038/nphoton.2017.93}
}

@article{Tait2017SiliconWeightBanks,
  author  = {Tait, Alexander N. and De Lima, Thomas F. and Zhou, Ellen and Wu, Anthony X. and Nahmias, Michael A. and Shastri, Bhavin J. and Prucnal, Paul R.},
  title   = {Neuromorphic Photonic Networks Using Silicon Photonic Weight Banks},
  journal = {Scientific Reports},
  volume  = {7},
  pages   = {7430},
  year    = {2017},
  doi     = {10.1038/s41598-017-07754-z}
}

@article{Honjo2021CIM100k,
  author  = {Honjo, T. and others},
  title   = {100,000-spin Coherent Ising Machine},
  journal = {Science Advances},
  volume  = {7},
  number  = {40},
  pages   = {eabh0952},
  year    = {2021},
  doi     = {10.1126/sciadv.abh0952}
}

@article{Berloff2017PolaritonXY,
  author  = {Berloff, Natalia G. and others},
  title   = {Realizing the Classical {XY} Hamiltonian in Polariton Condensates},
  journal = {Nature Materials},
  volume  = {16},
  pages   = {1120--1126},
  year    = {2017},
  doi     = {10.1038/nmat4971}
}

@misc{Fedus2022SwitchTransformers,
  author        = {Fedus, William and Zoph, Barret and Shazeer, Noam},
  title         = {Switch Transformers: Scaling to Trillion Parameter Models with Simple and Efficient Sparsity},
  howpublished  = {arXiv},
  year          = {2021},
  eprint        = {2101.03961},
  archivePrefix = {arXiv},
  primaryClass  = {cs.LG},
  url           = {https://arxiv.org/abs/2101.03961}
}

@article{wulf1995hitting,
  title={Hitting the memory wall: Implications of the obvious},
  author={Wulf, Wm A and McKee, Sally A},
  journal={ACM SIGARCH computer architecture news},
  volume={23},
  number={1},
  pages={20--24},
  year={1995},
  publisher={ACM New York, NY, USA}
}

@inproceedings{horowitz20141,
  title={1.1 computing's energy problem (and what we can do about it)},
  author={Horowitz, Mark},
  booktitle={2014 IEEE international solid-state circuits conference digest of technical papers (ISSCC)},
  pages={10--14},
  year={2014},
  organization={IEEE}
}

@article{hennessy2019new,
  title={A new golden age for computer architecture},
  author={Hennessy, John L and Patterson, David A},
  journal={Communications of the ACM},
  volume={62},
  number={2},
  pages={48--60},
  year={2019},
  publisher={ACM New York, NY, USA}
}

@misc{SM,
   title  = {See Supplemental Material},
   note   = {at
             [\url{https://www.damtp.cam.ac.uk/user/ngb23/publications/SI_AKM.pdf}]
             for soft-spin dynamics, binary comparator details, ridge
             regularization, training derivations, experiment
             specification, {$\sigma_{\min}$} scaling analysis, and
             reproducible code}
}

@article{Jaeger2004,
  author  = {Jaeger, Herbert and Haas, Harald},
  title   = {Harnessing nonlinearity: Predicting chaotic systems and
             saving energy in wireless communication},
  journal = {Science},
  volume  = {304},
  pages   = {78--80},
  year    = {2004}
}

@article{ramsauer2021,
  title={Hopfield networks is all you need},
  author={Ramsauer, Hubert and Sch{\"a}fl, Bernhard and Lehner, Johannes and Seidl, Philipp and Widrich, Michael and Adler, Thomas and Gruber, Lukas and Holzleitner, Markus and Pavlovi{\'c}, Milena and Sandve, Geir Kjetil and others},
  journal={arXiv preprint arXiv:2008.02217},
  year={2020}
}

@article{Dambre2012IPC,
  author  = {Dambre, Joni and Verstraeten, David and Schrauwen, Benjamin and Massar, Serge},
  title   = {Information Processing Capacity of Dynamical Systems},
  journal = {Scientific Reports},
  volume  = {2},
  pages   = {514},
  year    = {2012},
  doi     = {10.1038/srep00514},
}

@article{gao2020physical,
  title={Physical unclonable functions},
  author={Gao, Yansong and Al-Sarawi, Said F and Abbott, Derek},
  journal={Nature Electronics},
  volume={3},
  number={2},
  pages={81--91},
  year={2020},
  publisher={Nature Publishing Group UK London},
  doi={10.1038/s41928-020-0372-5}
}

@article{Pappu2002PUF,
  author  = {Pappu, Ravikanth and Recht, Ben and Taylor, Jason and Gershenfeld, Neil},
  title   = {Physical One-Way Functions},
  journal = {Science},
  volume  = {297},
  pages   = {2026--2030},
  year    = {2002},
  doi     = {10.1126/science.1074376},
}

@article{Oliver2015Consistency,
  author  = {Oliver, Neus and J{\"u}ngling, Thomas and Fischer, Ingo},
  title   = {Consistency Properties of a Chaotic Semiconductor Laser Driven by Optical Feedback},
  journal = {Physical Review Letters},
  volume  = {114},
  pages   = {123902},
  year    = {2015},
  doi     = {10.1103/PhysRevLett.114.123902},
}

@article{Ohadi2015SpinBifurcation,
  author  = {Ohadi, H. and Dreismann, A. and Rubo, Y. G. and Pinsker, F. and del Valle-Inclan Redondo, Y. and Tsintzos, S. I. and Hatzopoulos, Z. and Savvidis, P. G. and Baumberg, J. J.},
  title   = {Spontaneous Spin Bifurcations and Ferromagnetic Phase Transitions in a Spinor Exciton-Polariton Condensate},
  journal = {Physical Review X},
  volume  = {5},
  pages   = {031002},
  year    = {2015},
  doi     = {10.1103/PhysRevX.5.031002},
}

@article{Pickup2018SpinFlips,
  author  = {del Valle-Inclan Redondo, Yago and Ohadi, Hamid and Rubo, Yuri G. and Beer, Orr and Ramsay, Andrew J. and Tsintzos, Symeon I. and Hatzopoulos, Zacharias and Savvidis, Pavlos G. and Baumberg, Jeremy J.},
  title   = {Stochastic spin flips in polariton condensates: nonlinear tuning from {GHz} to sub-{Hz}},
  journal = {New Journal of Physics},
  volume  = {20},
  pages   = {075008},
  year    = {2018},
  doi     = {10.1088/1367-2630/aad377},
}

@article{Veraldi2025PRL,
  author    = {D. Veraldi and D. Pierangeli and S. Gentilini
               and M. {Calvanese Strinati} and J. Sakellariou
               and J. S. Cummins and A. Kamaletdinov and M. Syed
               and R. Z. Wang and N. G. Berloff
               and D. Karanikolopoulos and P. G. Savvidis
               and C. Conti},
  title     = {Fully Programmable Spatial Photonic {Ising} Machine
               by Focal Plane Division},
  journal   = {Physical Review Letters},
  volume    = {134},
  number    = {6},
  pages     = {063802},
  year      = {2025},
  doi       = {10.1103/PhysRevLett.134.063802}
}

@article{karanikolopoulos2026,
  title={High-Fidelity Spatial Photonic Ising Machines via Precise Wavefront Shaping},
  author={Karanikolopoulos, D and Karavelas, PS and Mouchliadis, L and Spiliotis, AK and Pitanios, NL and Gentilini, S and Veraldi, D and Charlesworth, P and Pierangeli, D and Sakellariou, J and others},
  journal={arXiv preprint arXiv:2602.13714},
  year={2026}
}

\end{document}